\documentclass[preprint,preprintnumbers,amsmath,amssymb,superscriptaddress]{revtex4-2}

\usepackage{fullpage,amsthm,amsmath,amsfonts,bm,times,color,bbm}
\usepackage{graphicx}
\usepackage[english]{babel}
\usepackage{mathtools,color}
\usepackage{lineno}
\usepackage{graphicx}% Include figure files
\usepackage{array}
\usepackage{float}
\usepackage{wrapfig}
\usepackage{hyperref}
\usepackage{romannum}
\usepackage{amsmath}
\usepackage{amssymb}
\usepackage{multirow}
\usepackage{makecell}
\usepackage[flushleft]{threeparttable}%% for tables

\usepackage{soul} %% for the comments of the text
\usepackage[table, xcdraw]{xcolor}
\usepackage{booktabs}
\usepackage{longtable}

\makeatother

\makeatletter

\begin{document}
	% \preprint{APS/123-QED}
	
	\pagenumbering{arabic}
	
	\title{Unraveling the mystery of tropical monsoon long-term prediction
 % I suggest the following title!
% Unraveling the Mystery of Tropical Monsoon Long-term Prediction
  }%
        % Optional Title:
        % \title{Percolation reveals the greening tipping point of the Tibet  plateau}%

	\author{Guanghao Ran}
        \affiliation{Department of Physics, Hong Kong Baptist University, Hong Kong, China}	
 
	\author{Jun Meng}
	\email{jun.meng.phy@gmail.com}
	\affiliation{National Key Laboratory of Earth System Numerical Modeling and Application, Institute of Atmospheric Physics, Chinese Academy of Sciences, Beijing 100029, China}
	%\affiliation{Potsdam Institute for Climate Impact Research, Potsdam 14412, Germany}

    \author{Jingfang Fan}%
	\email{jingfang@bnu.edu.cn}
	\affiliation{School of Systems Science/Institute of Non-equilibrium Systems, Beijing Normal University,  Beijing 100875, China}
	\affiliation{Potsdam Institute for Climate Impact Research, Potsdam 14412, Germany}
	
\begin{abstract}
Tropical monsoons play a critical role in shaping regional and global climate systems, with profound ecological and socio-economic impacts. However, their long-term prediction remains challenging due to the complex interplay of regional dynamics, global climate drivers, and large-scale teleconnections. Here, we introduce a unified network-based framework for predicting monsoon precipitation across diverse tropical regions. By leveraging global 2-meter air temperature fields,  this approach captures large-scale climate teleconnections, such as the El Niño-Southern Oscillation (ENSO) and Rossby waves, enabling accurate forecasts for four key monsoon systems: the South American, East Asian, African, and Indian monsoons. Our framework achieves remarkable forecasting accuracy with lead times of 4–10 months, outperforming traditional systems such as SEAS5 and CFSv2. Beyond its predictive capabilities, the framework offers flexibility for application to other regions and climate phenomena, advancing our understanding of global climate dynamics. These findings have far-reaching implications for disaster preparedness, resource management, and sustainable development.
%Tropical monsoons exert a significant influence on the climatic and ecological systems of tropical regions, with impacts extending globally. However, the complexity of regional to global atmosphere-ocean interactions poses a great challenge in long-term monsoon prediction. In our study, we present a complex network-based framework for monsoon prediction that utilizes the global 2-meter air temperature field. We demonstrate that specific characteristics of the climate network can effectively serve as long-term predictors for forecasting four typical tropical monsoons. Notably, our framework demonstrates significant forecasting skill with a lead time of 4-10 months, surpassing the performance of traditional numerical prediction methods such as SEAS5 and CFSv2. The superior performance of our network-based framework may be attributed to its ability to capture teleconnections driven by large-scale climate phenomena such as the El Niño-Southern Oscillation (ENSO) and Rossby waves within the climate system. This network-based framework can potentially be applied to other regions and climate phenomena, enhancing our understanding and prediction capabilities for global climate patterns.

\end{abstract}
\date{\today}
\maketitle
% \tableofcontents

\section{Introduction}
Tropical monsoon regions, concentrated within  $30^\circ$ latitude of the equator, are marked by abundant precipitation and exceptional ecological diversity. Monsoon precipitation is vital for maintaining regional ecosystems \cite{fu1999variation}, supporting agriculture \cite{wang2006asian, gadgil2006indian}, and water resource management \cite{balek1983hydrology}.  Beyond their local significance, tropical monsoons influence global atmospheric circulation, water cycling \cite{worden2007importance}, and energy balance \cite{trenberth2000global}.

However, accurately forecasting monsoon precipitation over long timescales remains a formidable challenge due to the complex interplay of initial conditions, external forcings, and internal variability \cite{zhou2011global}.
For instance, small errors in initial conditions can amplify over time, making accurate long-term forecasts difficult \cite{slingo2011uncertainty}. Additionally, uncertainties in future external forcings, such as greenhouse gas emissions and volcanic activity, further complicate predictions \cite{collins2013long, wang2015rethinking}. The strong natural variability of monsoons adds another layer of difficulty, posing significant challenges for seasonal and decadal-scale forecasting \cite{kajtar2017tropical}.

Despite progress, climate models struggle to capture the nonlinear processes and complex interactions that define the climate system, limiting their long-term forecasting accuracy \cite{sperber2013asian}. Furthermore, these models require extensive observational data and sophisticated parameterizations, complicating their use in regions with limited historical records \cite{fan2012improving, shi2021significant}. Recently, AI-based methods have enabled the development of machine learning weather prediction (MLWP) systems, such as FourCastNet \cite{pathak2022fourcastnet}, GraphCast \cite{lam2023learning}, Pangu-Weather \cite{bi2023accurate}, NowcastNet \cite{zhang2023skilful}, and GenCast \cite{price2023gencast}. However, these systems excel at short- to medium-term (up to 15 days) predictions and are currently ill-suited for addressing the complexities of long-term and seasonal-scale precipitation forecasting.

To address these challenges, network theory offers a robust approach for analyzing the dynamic and structural properties of complex systems \cite{watts1998collective, albert2002statistical, newman2018networks, cohen2010complex}. Its application in climate science \cite{tsonis2004architecture, yamasaki2008climate,boers_prediction_2014, boers_complex_2019} has revealed hidden connections and teleconnections within the global climate system, providing a novel basis for prediction \cite{feng_are_2014,dijkstra2019networks,ludescher2021network}. By leveraging network-derived metrics, researchers have successfully predicted phenomena like El Niño events \cite{ludescher2013improved, meng2020complexity} and monsoonal rainfall \cite{boers_prediction_2014,fan2022network}, demonstrating the approach's feasibility and reliability.

We build on these advancements by introducing a climate network-based framework for predicting monsoon precipitation in four key tropical regions, each representing a major monsoon system: the South American monsoon (Eastern Amazon rainforest, EAR), East Asian monsoon (Hainan Island, HI), African monsoon (Sahel region of East Africa, SEA), and Indian monsoon (Central India, CI). Our framework achieves high accuracy and extended lead times, with correlation coefficients ($r$-values) exceeding 0.6 and mean absolute percentage errors (MAPE) below 16\%, across lead times of 4–10 months. Through a comparative analysis with SEAS5 \cite{johnson2019seas5} and CFSv2 \cite{saha2014ncep}, our method demonstrates superior performance in prediction accuracy and lead time.

\section{Results}

\subsection{Mining network predictors \label{predictors}}

%Monsoons, characterized by a significant shift in wind patterns, act as heralds of heavy rainfall. These monsoonal winds play a pivotal role in determining the onset and duration of the rainy season, especially in regions prone to monsoons. The rainy season, marked by a substantial increase in precipitation compared to other times of the year, exhibits considerable variability in terms of its timing, duration, and intensity. The southwest monsoon, which assumes a prominent role as the primary rainy season in various parts of India, Bangladesh, and neighboring countries, typically spanning from June to September. In regions such as China and Japan, a separate and distinguishable rainy season, often referred to as the "plum rain" season, graces the early summer months, primarily occurring in June and July. On a different note, the Amazon rainforest generally experiences its rainy season during the boreal autumn, which aligns with the southern hemisphere's summer months, typically from November to April. To provide a streamlined categorization and encompass the entirety of the rainy season, we have defined the rainy season in the northern hemisphere (NH) as extending from May to October, while the rainy season in the southern hemisphere (SH) spans from November to April.

The rainy season, characterized by significant increases in precipitation, varies widely in timing, duration, and intensity across different regions. For this study, we focused on four regions within the tropical monsoon belt: the Eastern Amazon Rainforest (EAR), Hainan Island (HI), the Sahel region of East Africa (SEA), and Central India (CI). These regions represent diverse monsoonal systems driven by distinct atmospheric and oceanic processes, making them ideal test cases for our predictive framework. To standardize the analysis, we defined the rainy season as May to October in the Northern Hemisphere and November to next year's April in the Southern Hemisphere, based on historical precipitation patterns.

In our endeavor to identify reliable predictors for rainy season precipitation, we constructed a series of climate networks spanning January 1950 to December 2022. The dataset used for this analysis comprises daily mean 2-meter air temperature data from NCEP-Reanalysis I, covering 6,242 global grid points. Each grid point was treated as a node within the network, and its in-degree and out-degree were computed based on correlations with other nodes (See Methods). For a given year \( y \), the in-degree of a node, denoted as \( k^{\text{in}+}_{i}(y) \) or \( k^{\text{in}-}_{i}(y) \), quantifies the influence it receives from other regions globally. Similarly, the out-degree, represented as \( k^{\text{out}+}_{i}(y) \) or \( k^{\text{out}-}_{i}(y) \), reflects the extent to which the node influences other regions. The positive or negative sign indicates whether these relationships are reinforcing (positive) or opposing (negative). For instance, a high \( k^{\text{out}+}_{i}(y) \) or \( k^{\text{out}-}_{i}(y) \) suggests that temperature fluctuations at node \( i \) during year \( y \) significantly impact other regions, either promoting similar changes (positive) or opposite changes (negative).

Our primary objective was to identify specific regions within the tropical monsoon belt where rainy season precipitation could be reliably predicted, alongside their corresponding network predictors. To achieve this, we analyzed network nodes, focusing on those whose in-degree or out-degree showed strong correlations with precipitation in targeted monsoon regions. To ensure the robustness of these predictors under changing climate conditions, we divided the analysis period into two segments: an initial phase (1950--1989) and a subsequent phase (1990--2022). Predictors were deemed reliable if their correlations with precipitation were statistically significant (\( p < 0.05 \)) in both periods and stronger during the latter phase, reflecting their adaptability to global warming trends.

\begin{figure}[hbpt]
\centering
    \includegraphics[width=0.95\linewidth]{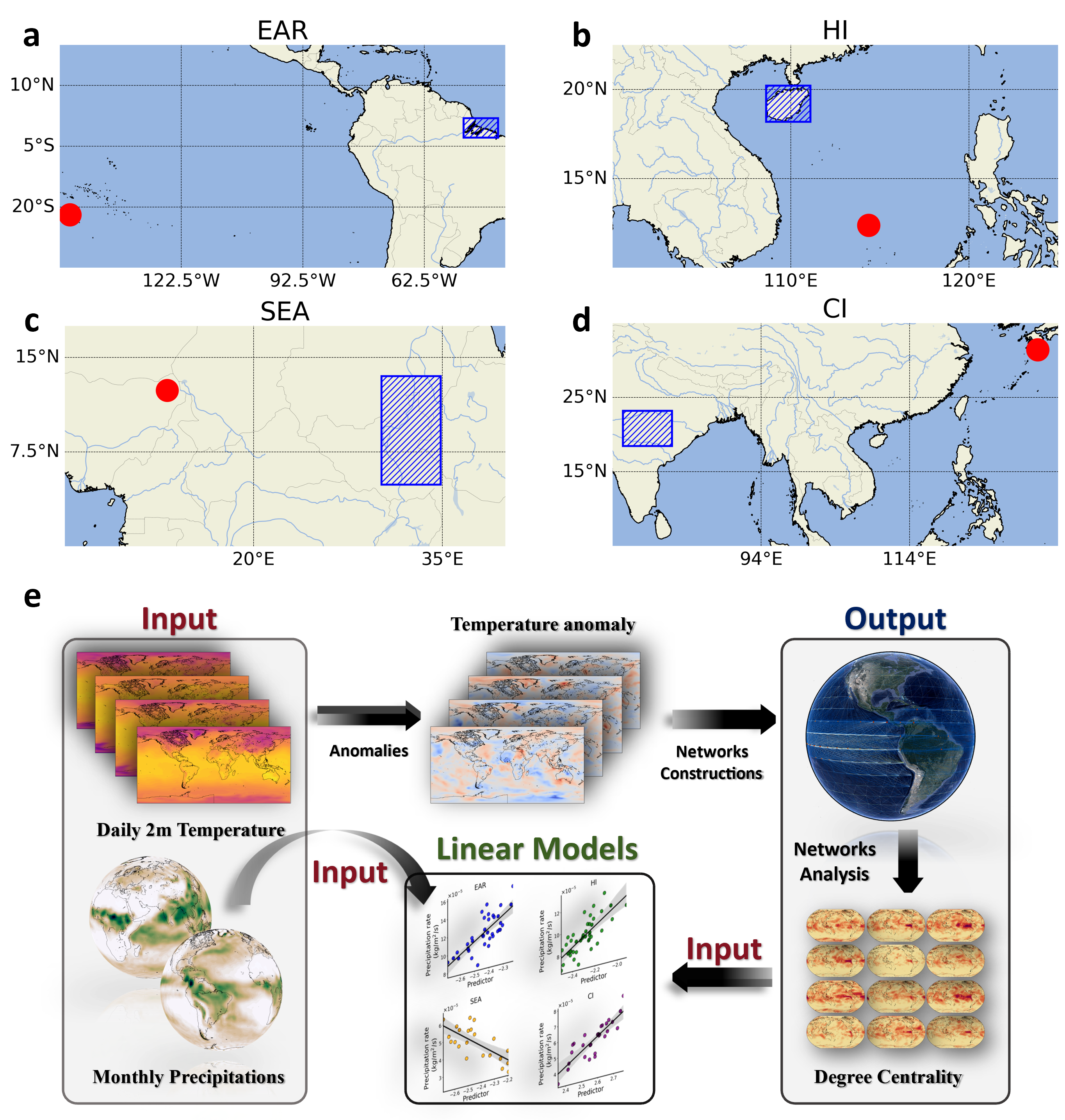}
    \caption{ { \bf Network-based prediction framework.} The red dots on the map indicate the locations of the identified predictors, while the blue boxes delineate the targeted forecast regions: (\textbf{a}) EAR, (\textbf{b}) HI, (\textbf{c}) SEA, and (\textbf{d}) CI. Panel (\textbf{e}) presents the proposed prediction framework, highlighting the integration of network-based metrics and statistical modeling for monsoon precipitation forecasting.}
    \label{map}%
\end{figure}

To further enhance reliability, we prioritized predictors exhibiting the strongest correlations during the recent 33-year period. Through a comprehensive global analysis, we identified specific nodes (highlighted in red in Figure.~\ref{map}) with in-degree or out-degree metrics that exhibited the most robust correlations with rainy season precipitation in four key monsoon regions, i.e. EAR, HI, SEA and CI, as shown in Fig.~S1.

\begin{figure}[hbpt]
	\centering
    \includegraphics[width=1\linewidth]{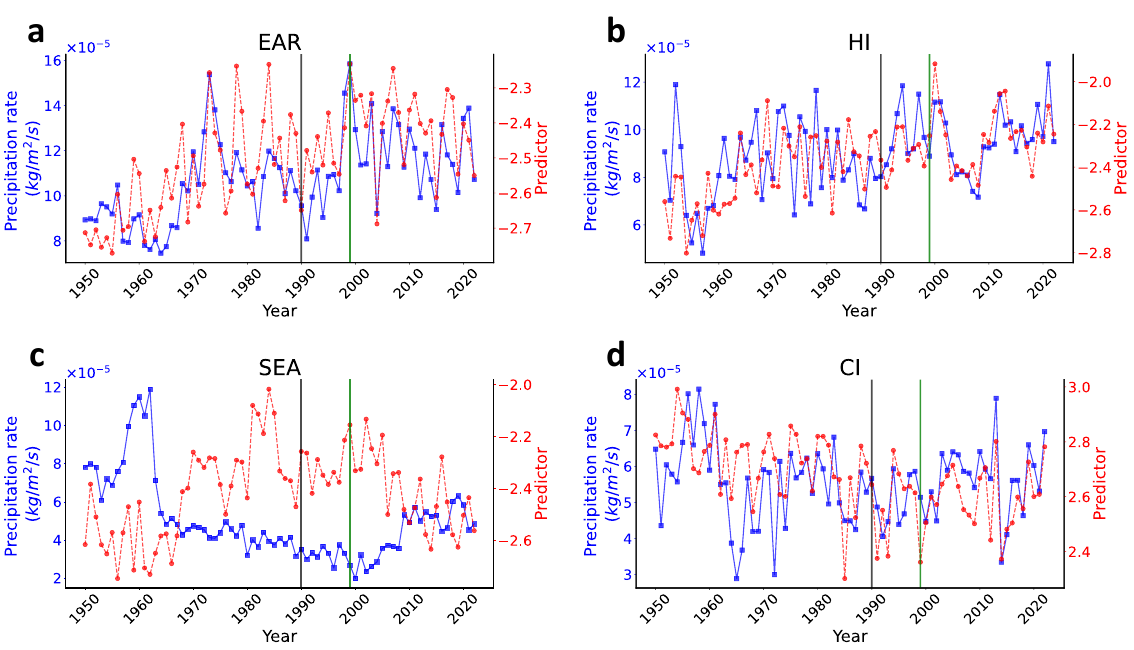}
   \caption{\textbf{A visualization of the linear relationship between network predictors and rainy season precipitation:} (\textbf{a}) EAR, (\textbf{b}) HI, (\textbf{c}) SEA, and (\textbf{d}) CI. The blue curves represent the time series of observed seasonal total precipitation rates in tropical monsoon regions, while the red curves represent their corresponding network predictors. The black and green vertical lines indicate the years 1990 and 1999, respectively, dividing the entire time span into three phases: the learning phase (1950--1989), the training phase (1990--1999), and the forecasting phase (2000--2022).}
    \label{linearity}
\end{figure}

These regions, illustrated as blue boxes in Figure.~\ref{map}, encompass EAR ($0.9524^\circ\text{S}$ to $0.9524^\circ\text{N}$, $46.8750^\circ\text{W}$ to $52.5000^\circ\text{W}$), HI ($18.0950^\circ\text{N}$ to $19.9997^\circ\text{N}$, $108.7500^\circ\text{E}$ to $110.6250^\circ\text{E}$), SEA ($6.6666^\circ\text{N}$ to $12.3808^\circ\text{N}$, $31.8750^\circ\text{E}$ to $33.7500^\circ\text{E}$), and CI ($19.9997^\circ\text{N}$ to $21.9044^\circ\text{N}$, $76.8750^\circ\text{E}$ to $80.6250^\circ\text{E}$). 
We designate the corresponding in-degree or out-degree of these nodes as predictors (as shown in Table.~\ref{table_predictors}) for rainy season precipitation in these four regions. The calculation of rainy season precipitation for these regions is based on their respective geographical locations (refer to the Methods section for details). An illustrative depiction of the linear relationship between network predictors and rainy season precipitation is presented in Fig.~\ref{linearity}, offering a clear visualization of their correlation.

\begin{table}[hbpt]
\caption{\textbf{Predictors' information for monsoon precipitation forecasting}}
\begin{tabular}{cccc}
\hline
{\color[HTML]{000000} \textbf{Monsoon region}} & {\color[HTML]{000000} \textbf{Rainy season}} & {\color[HTML]{000000} \textbf{Predictor's location}}                                                                                                                                             & {\color[HTML]{000000} \textbf{Predictor}} \\ \hline
EAR                     & Nov. to Apr.                                 & ($21.9044^\circ$S, $150.0000^\circ$W)                                                                                                                                                               & $k^{out-}_{4279}$                                \\
HI                                 & May. to Oct.                                 & ($12.3808^\circ$N, $114.3750^\circ$E)                                           & $k^{out-}_{2500}$                                \\
SEA                        & May. to Oct.                                 & ($12.3808^\circ$N, $13.1250^\circ$E)                                            & $k^{out-}_{2446}$                                 \\

CI                                 & May. to Oct.                                 & ($31.4281^\circ$N, $131.2500^\circ$E)                                                                                                                                                               & $k^{out+}_{1549}$                                \\
 \hline
\end{tabular}
\label{table_predictors}
\end{table}

\subsection{Forecasting the monsoon precipitation}

We identified a strong linear relationship between specific predictors and precipitation rates in each of the targeted regions (illustrated in Fig.~\ref{linearity}). Leveraging this relationship, we developed reliable forecasting algorithms using a linear regression model. However, given the non-stationary nature of the climate system, where dynamics change over time, we assumed that the linear relationships would also evolve dynamically. To address this, we partitioned the most recent 33 years into two phases and fine-tuned our model by determining the optimal window length, $L_{opt}$, during the training phase (1990–1999). Rather than relying on all historical data since 1950, the model employs a moving window of data with a fixed length $L_{opt}$. Subsequently, the forecasting phase (2000–2022) is used to perform actual predictions and evaluate the algorithm's performance. This adaptive approach allows the model to adjust to recent trends and variability, mitigating the influence of outdated or fluctuating data.

\begin{figure}[hbpt]
\centering
    \includegraphics[width=.9\linewidth]{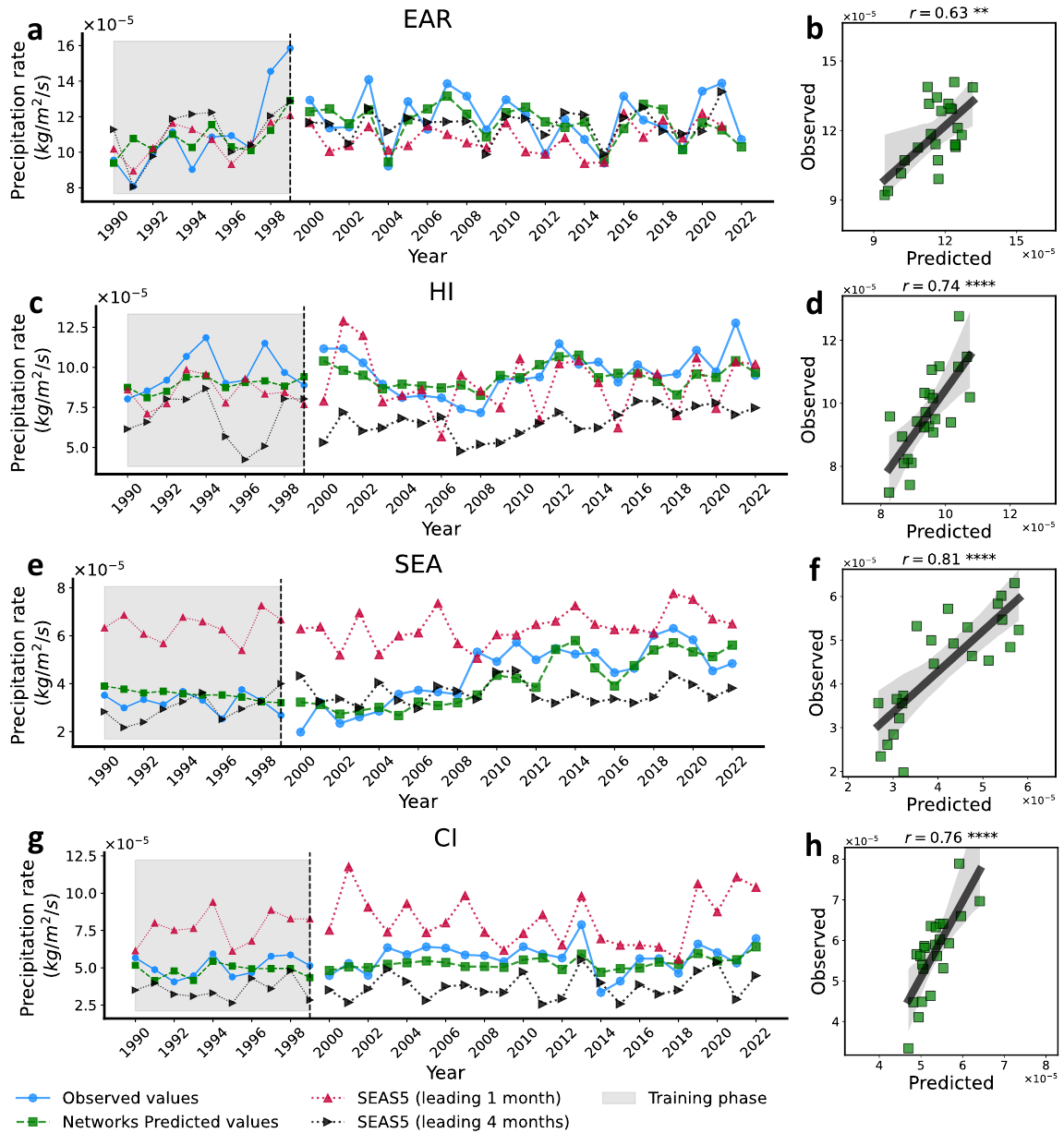}
    \caption{\textbf{Prediction results.} 
     Panels \textbf{(a), (c), (e)}, and \textbf{(g)} display time series of predicted values (green dashed lines) and observed values (blue solid lines). The SEAS5 hindcast values with lead times of 1 month and 4 months are shown as pink and black dotted lines, respectively. The grey shaded area indicates the training phase. Panels \textbf{(b), (d), (f)}, and \textbf{(h)} display the scatter plot of predicted values versus observed values for the period from 2000 to 2022. The black solid line represents the linear regression line, showing the overall trend. The shaded area represents the 99\% confidence interval of the regression line. The markers $^{\ast\ast\ast\ast}$ and $^{\ast\ast}$ indicate $p < 10^{-4}$ and $10^{-2}$, respectively.}
    \label{fig_prediction_1}%
\end{figure}
%We have identified a strong linear relationship between certain predictor and the precipitation rate in each of the targeted regions (as illustrated in Fig.~\ref{linearity}). Leveraging this relationship, we developed reliable forecasting algorithms using a linear regression model. However, due to the non-stationary nature of the climate system, which changes over time, we assume that the linear relationships may also change dynamically. To address this, we partitioned the most recent 33 years into two phases and fine-tuned our model by determining the optimal window length $L_{opt}$ for our linear regression model based on the training phase, which spans from 1990 to 1999. Instead of relying on all past data since 1950, the model utilizes a moving window of data with a fixed length $L_{opt}$. Following this, the forecasting phase (2000-2022) is dedicated to actual forecasting and testing the algorithm's performance. This adaptive approach enables the model to adjust to recent trends and changes, thereby mitigating the influence of data fluctuations to a certain extent.

To assess the performance of our forecasting framework, we employed three key metrics: Root Mean Square Error (RMSE) \cite{willmott2005advantages}, Mean Absolute Percentage Error (MAPE) \cite{de2016mean}, and Pearson Correlation Coefficient (PCC) \cite{cohen2009pearson}. RMSE measures the average magnitude of prediction errors, highlighting significant deviations. MAPE expresses errors as a percentage, enabling easy comparison across datasets. PCC evaluates the linear correlation between predicted and observed values, indicating trend consistency. We systematically optimized these metrics to determine the best parameter set for our prediction framework by minimizing RMSE and MAPE while maximizing PCC. Specifically, 
we defined an evaluation criterion $\lambda$ and selected the value of $L$ that maximized $\lambda$ as the optimal window length:
\begin{equation}
    L_{opt} = \arg\max_{L} \lambda = \arg\max_{L}\left( \frac{PCC_{norm}}{RMSE_{norm}\cdot MAPE_{norm}} \right),
    \label{EQ:1}
\end{equation}
where PCC$_{norm}$, RMSE$_{norm}$ and MAPE$_{norm}$ represent the normalized PCC, RMSE and MAPE respectively. Our investigation revealed that employing a fixed $L$ of 40 years, corresponding to the 40-year period preceding the predicted year $y$, resulted in poor performance in terms of RMSE and MAPE. This finding suggests the presence of significant systematic errors (see Appendix Fig.~S2 for details).

\begin{longtable}{l *{9}{c}}
\caption{\textbf{A comparison of the forecast performance among SEAS5, CFSv2, and the network-based prediction method.}}
\label{table_prediction} \\
\toprule
 &
  \multicolumn{3}{c}{\textbf{PCC}} &
  \multicolumn{3}{c}{\textbf{MAPE (\%)}} &
  \multicolumn{3}{c}{\textbf{RMSE ($\mathbf{\times 10^{-5}kg/m^2/s}$)}} \\
\cmidrule(lr){2-4} \cmidrule(lr){5-7} \cmidrule(lr){8-10}
\textbf{Regions} &
  \textit{\textbf{SEAS5}} &
  \textit{\textbf{CFSv2}} &
  \textit{\textbf{Networks}} &
  \textit{\textbf{SEAS5}} &
  \textit{\textbf{CFSv2}} &
  \textit{\textbf{Networks}} &
  \textit{\textbf{SEAS5}} &
  \textit{\textbf{CFSv2}} &
  \textit{\textbf{Networks}} \\
\midrule
\endfirsthead
\toprule
 &
  \multicolumn{3}{c}{\textbf{PCC}} &
  \multicolumn{3}{c}{\textbf{MAPE (\%)}} &
  \multicolumn{3}{c}{\textbf{RMSE ($\mathbf{\times 10^{-5}kg/m^2/s}$)}} \\
\cmidrule(lr){2-4} \cmidrule(lr){5-7} \cmidrule(lr){8-10}
\textbf{Regions} &
  \textit{\textbf{SEAS5}} &
  \textit{\textbf{CFSv2}} &
  \textit{\textbf{Networks}} &
  \textit{\textbf{SEAS5}} &
  \textit{\textbf{CFSv2}} &
  \textit{\textbf{Networks}} &
  \textit{\textbf{SEAS5}} &
  \textit{\textbf{CFSv2}} &
  \textit{\textbf{Networks}} \\
\midrule
\endhead
\bottomrule

\endfoot
\endlastfoot
\textit{\textbf{EAR}} &
  \textbf{0.65} &
  0.67 &
  0.63 \textbf{(0.77)} &
  10.48 &
  35.7 &
  \textbf{7.57 (6.72)} &
  1.55 &
  4.45 &
  \textbf{1.13 (0.93)} \\
\textit{\textbf{HI}} &
  0.47 &
  0.29 &
  \textbf{0.74 (0.82)} &
  15.81 &
  43.91 &
  \textbf{8.24 (7.96)} &
  1.77 &
  4.43 &
  \textbf{0.95 (0.81)} \\
\textit{\textbf{SEA}} &
  0.36 &
  0.44 &
  \textbf{0.81 (0.77)} &
  58.10 &
  27.27 &
  \textbf{15.31 (17.20)} &
  2.28 &
  1.11 &
  \textbf{0.76 (0.82)} \\
\textit{\textbf{CI}} &
  0.36 &
  0.53 &
  \textbf{0.76 (0.84)} &
  47.63 &
  14.89 &
  \textbf{12.65 (14.27)} &
  3.02 &
  1.01 &
  \textbf{0.80 (0.89)} \\
\bottomrule
\multicolumn{10}{@{}l}{\footnotesize \textit{Note:} SEAS5 (2000--2022), CFSv2 (2000--2016). Bracketed values: network-based method (2000--2016).}
\end{longtable}

Consequently, we determined the optimal learning set lengths $L_{opt}$ for four monsoon regions as follows: 37 years for EAR, 40 years for HI,  11 years for SEA, and 33 years for CI (see Fig.~S3 for details). Using these optimal learning set lengths, we forecasted rainy season total precipitation for the respective regions from 2000 to 2022, yielding favorable results. The forecasts achieved: RMSE values below $1.2\times10^{-5}kg/m^2/s$, MAPE values below 16\%, and PCC values higher than $0.6$ (refer to TABLE. \ref{table_prediction}). %Notably, the EAR region had the smallest $MAPE$ at 7.57\% (see Figure\ref{fig_prediction_1}{\bf a}). The SEA region exhibited the highest $PCC$ of $0.81$ and the lowest $RMSE$ of $0.76\times10^{-5}kg/m^2/s$ (see Figure \ref{fig_prediction_1}{\bf c}).

This prediction model demonstrates a significant advantage over alternative forecasting methods, particularly in terms of lead time. For SEA, CI, and HI, our model enables predictions with a lead time of up to \textit{4 months,} while for EAR, the lead time extends to \textit{10 months}. Such extended forecasting capabilities are crucial for societal, economic, and environmental applications, offering improved responses to climate-related disasters, enhanced resource management, and additional temporal windows for informed decision-making and planning.

% This prediction model boasts a significant advantage over alternative forecasting methods, particularly in terms of lead time. For SEA, CI, and HI, our model enables predictions up to 4 months in advance, whereas for EAR, we extend the lead time to 10 months. Such advanced forecasting capabilities hold paramount importance for societal, economic, and environmental aspects, enabling improved responses to climate-related disasters, enhanced resource management, and providing additional information and temporal windows for decision-making and planning. %Furthermore, our climate network-based prediction model, underpinned by robust physical mechanisms, offers superior interpretability ( \cite{fan2022network, boers2014prediction, meng2020complexity}) compared to contemporary AI-based prediction approaches like {\it Pangu-Weather}  \cite{bi2023accurate}.

\subsection{Comparative study: Seasonal Forecast Systems vs. Network Based Prediction Method}

We conducted a comparative analysis to evaluate the predictive performance of our proposed network-based forecasting method against two established seasonal climate prediction systems: the Seasonal Ensemble Prediction System 5 (SEAS5) developed by the European Centre for Medium-Range Weather Forecasts (ECMWF) and the Climate Forecast System version 2 (CFSv2) developed by the National Oceanic and Atmospheric Administration (NOAA). Both SEAS5 and CFSv2 are known to provide a reasonable fit for the observed spatial distribution of precipitation, as shown in Figs. S3 and S4. It is noteworthy that the hindcast data for SEAS5 is only available up to 2022, while the hindcast data for CFSv2 has gaps during 1990-1992 and 2017-2018. Therefore, only the period from 2000 to 2016 was used to calculate the hindcast performance of CFSv2. Numerous studies have evaluated the performance of SEAS5 and CFSv2 in tropical monsoon regions \cite{ferreira2022evaluation, chevuturi2021forecast, lang2014evaluating}. For instance, Ferreira et al. \cite{ferreira2022evaluation} demonstrated that SEAS5 effectively captures extreme precipitation anomalies in northern South America and northeastern Brazil, despite some spatial heterogeneity. Chevuturi et al. \cite{chevuturi2021forecast} reported a PCC of approximately 0.33 between SEAS5 hindcast data and observed data for the Indian monsoon. Similarly, Lang et al. \cite{lang2014evaluating} noted that CFSv2’s forecast skill scores for rainy season precipitation in 17 regions across China, including the Pearl River Basin (HI), generally remained below 0.3.

To compare the performance of SEAS5, CFSv2, and our network-based method, we evaluated their predictive capabilities for four tropical monsoon regions—EAR, HI, SEA, and CI-using hindcast data. SEAS5 and CFSv2 hindcast data covered the periods 1990–2022 and 1993–2016, respectively, with lead times of 1 and 4 months (Figs.~\ref{fig_prediction_1} and S5). SEAS5 utilizes 25 ensemble members \cite{johnson2019seas5}, while CFSv2 employs 24 ensembles \cite{saha2014ncep}. We compared these systems’ predictive performance against our network-based method using key metrics (e.g., RMSE, MAPE, and PCC). Table.~\ref{table_prediction} summarizes their performance for lead times of 1 month (refer to Table. S1 for lead-4 month results).

\begin{figure}[hbpt]
\centering
    \includegraphics[width=1\linewidth]{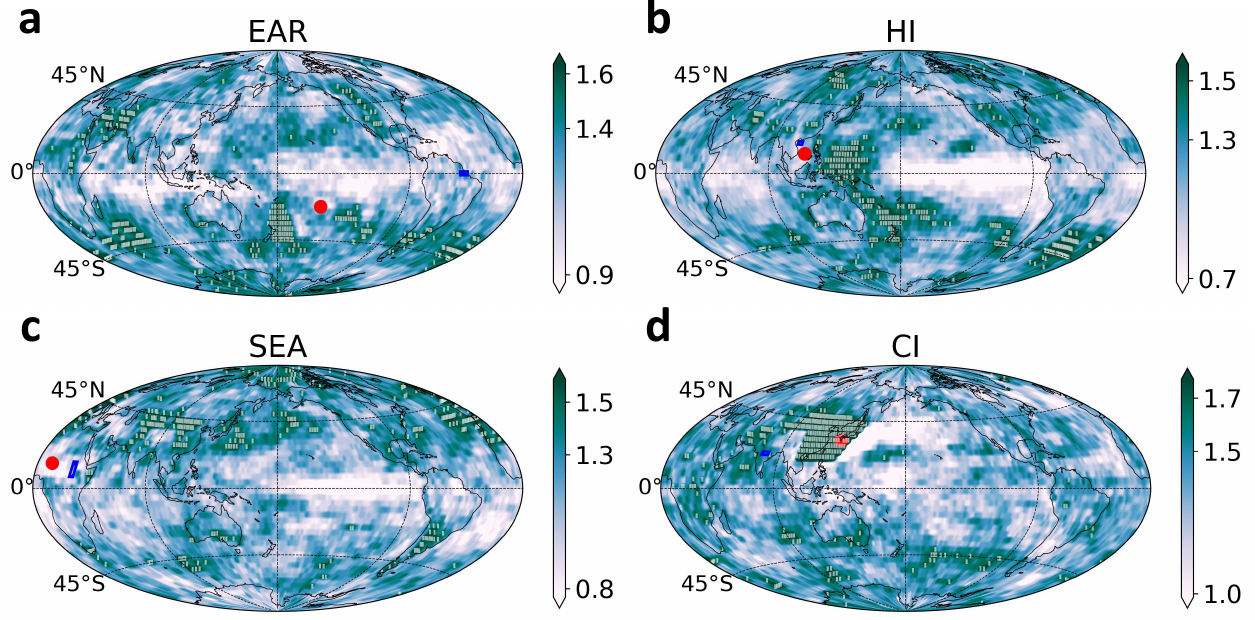}
   \caption{\textbf{Spatial distribution of the annually averaged absolute link strengths that comprise the predictor indices.} Red dots represent the locations of the predictors, while white cubes indicate the top 5\% of link strength values.}
    \label{predictors_degree}%
\end{figure}

SEAS5 and CFSv2 exhibited their best performance in the EAR region, achieving a maximum PCC of 0.67 at a lead time of 1 month. However, our network-based method matched their PCC (0.63) at a much longer lead time of 10 months, while also delivering significantly lower MAPE and RMSE values (Table.~\ref{table_prediction}). In contrast, SEAS5 and CFSv2 showed poorer performance in the other three regions, with PCC values ranging from 0.26 to 0.47 and higher errors compared to our method, as illustrated in Fig.~\ref{predictors_degree} and Fig. S4. These results are consistent with findings from prior studies \cite{ferreira2022evaluation, chevuturi2021forecast, lang2014evaluating}, further underscoring the advantages of our forecasting method.

Beyond its superior predictive performance, the network-based approach provides valuable insights into the physical mechanisms driving climate variability, addressing the opacity of AI models that often act as "black boxes". By analyzing the annual averaged absolute weights of network links connected to predictors, we identified regions with significant influences on predictor dynamics. These insights also helped relate network indices’ predictability to underlying climate mechanisms.
Key regions influencing predictors for the EAR and SEA (depicted in Fig.~\ref{predictors_degree} \textbf{a, c}) align with the Rossby wave train, consistent with the findings of Gelbrecht et al. \cite{gelbrecht2018phase}, which highlight the pivotal role of Rossby waves in modulating South American precipitation patterns. The propagation of Rossby waves across the atmosphere establishes teleconnections that influence precipitation variability in both the Eastern Amazon Rainforest and the Sahel region. For the HI region, influential predictors are primarily located in the eastern Pacific (Fig.~\ref{predictors_degree}\textbf{b}), a region strongly governed by the ENSO. ENSO's global impacts on atmospheric circulation and weather patterns, including its modulation of East Asian monsoons, have been extensively documented \cite{changnon2000nino, ronghui1989influence, yuan20172016}. The connection between ENSO phases and HI predictors underscores the critical role of large-scale ocean-atmosphere interactions in shaping rainfall patterns over East Asia. In the CI region, predictors cluster predominantly in East Asia, illustrating a teleconnection with the East Asian monsoon (Fig.~\ref{predictors_degree}\textbf{d}). This is consistent with prior studies, such as Kripalani et al. \cite{kripalani2001monsoon}, which demonstrated an out-of-phase relationship between monsoon precipitation in southern Japan and India. This teleconnection reflects the interplay between subtropical and tropical circulation systems, contributing to the variability of Indian monsoon rainfall.

These findings validate the network-based method as a powerful tool for understanding and forecasting monsoonal precipitation. By identifying and quantifying the influence of key teleconnections, this approach not only enhances predictive accuracy but also provides critical insights into the underlying physical mechanisms driving climate variability.

\section{Discussion}
In this study, we introduced a novel network-based framework for predicting monsoon precipitation across four tropical regions, each representing distinct monsoon systems: the South American monsoon (EAR), East Asian monsoon (HI), African monsoon (SEA), and Indian monsoon (CI). Our method outperformed established seasonal forecast systems like SEAS5 and CFSv2, achieving superior predictive accuracy with lead times of up to 10 months. Specifically, our framework delivered higher PCCs and notably lower MAPEs and RMSEs, particularly in the EAR and SEA regions. This approach provides several advantages. First, it dynamically adjusts predictors to account for evolving climate conditions, ensuring robustness in a non-stationary climate system and enabling more reliable long-term forecasts. Second, the method enhances interpretability by identifying influential predictor regions and their climatic linkages, such as the roles of Rossby waves and the ENSO. This transparency addresses a key limitation of AI-based models, which often function as "black boxes". Finally, the framework achieves extended lead times of up to 10 months, significantly outperforming SEAS5 and CFSv2. These capabilities are crucial for disaster preparedness, agricultural planning, and water resource management in monsoon-dependent regions.

The ability to forecast monsoon precipitation with high accuracy and extended lead times has profound implications. Reliable long-term forecasts enable early warning systems for extreme precipitation events, reducing vulnerability in regions heavily reliant on monsoonal rainfall. Accurate predictions also support sustainable resource management, particularly in agriculture and water systems. Furthermore, by linking network predictors to established teleconnections such as Rossby waves and ENSO, this framework enhances understanding of the complex interactions within the climate system, advancing the field of climate science.

While our framework demonstrates strong predictive capabilities, opportunities exist to further enhance its accuracy and applicability. Incorporating additional climatic variables—such as sea surface temperatures, atmospheric pressure patterns, and soil moisture—could provide a more comprehensive representation of monsoon dynamics and potentially refine prediction accuracy. Although the study focuses on four tropical monsoon regions, the framework's adaptability suggests that its application can be extended to other climatic phenomena, including extratropical systems and drought prediction, to further validate its robustness across diverse environments. Moreover, integrating machine learning techniques within the network-based framework \cite{nooteboom_using_2018,petersik_probabilistic_2020} holds promise for combining the interpretability of network analysis with the predictive power of AI, potentially offering a synergistic approach to advancing climate prediction.

By addressing the limitations of existing methods and offering deep insights into the physical mechanisms governing monsoonal precipitation, our framework represents a significant advancement in climate prediction. Its demonstrated ability to achieve extended lead times with high accuracy establishes it as a valuable tool for climate resilience planning, resource management, and disaster preparedness in vulnerable regions. Furthermore, the framework’s flexibility and interpretability pave the way for broader applications to other climate-related challenges, contributing to a more sustainable and informed global society.
%\clearpage
%\newpage

\section{Methods}
\subsection{Data} \label{datasection}
\subsubsection{Data availability}
The data come from the following sources: $\rm (1)$ daily means 2m temperature (1948-2021) from NCEP-NCAR reanalysis I, which have a spatial (longitudinal and latitudinal) resolution of 1.875°$\times$1.875° in Gaussian grid (\url{https://downloads.psl.noaa.gov/Datasets/ncep.reanalysis/Dailies/surface_gauss/}). This data field is chosen for two primary reasons: $\rm (a)$ it is highly correlated to the sea surface temperature(SST) and captures the atmosphere-ocean-land interaction processes, and $\rm (b)$ it can be updated on time, allowing us to make regular predictions; $\rm (2)$ monthly precipitation rate (1950-2022) \url{https://downloads.psl.noaa.gov/Datasets/ncep.reanalysis/Monthlies/surface_gauss/}) from NCEP-NCAR reanalysis I; $\rm (3)$ hindcast monthly means total precipitation from SEAS5 and CFSv2, and $\rm (5)$ monthly means total precipitation (1990-2022) from ERA5 are openly available from the Copernicus Climate Change Service (C3S) Climate DataStore, \url{https://cds.climate.copernicus.eu/#!/home}.

\subsubsection{Data pre-process}
For each node $i$ (i.e., longitude-latitude grid point), we calculate daily atmospheric temperature anomalies $T_{i}(t)$ (actual temperature value minus the climatological average and divided by the climatological standard deviation) for each calendar day. Specifically, given an actual record $\widetilde{T}^{y}(d)$, where $y$ is the year (from 1948 to 2021), and $d$ stands for the day (from 1 to 365), the anomalies record is defined as
\begin{equation}
    T^{y}(d) = \frac{\widetilde{T}^{y}(d)-\overline{\widetilde{T}^{y}(d)}}{\sigma(\widetilde{T}^{y}(d))},
\end{equation}
where
\begin{equation}
    \overline{\widetilde{T}^{y}(d)}=\left\{
    \begin{array}{cc}
        \frac{1}{y-1948} \sum^{y-1}_{k=1949}\widetilde{T}^{k}_{i}(d)   &  y\geq 1953 \\
        \frac{1}{4} \sum^{1952}_{k=1949}\widetilde{T}^{k}_{i}(d)   & y < 1953
    \end{array}
    \right.,
\end{equation}
similarly, $\sigma(\widetilde{T}^{y}(d))$ is calculated in the same segments. Leap days were excluded to maintain consistency in the calendar year duration and simplify the analysis.

To analyze regional precipitation patterns, we selected a varying number of nodes ($N_{EAR}, \\N_{HI}, N_{SEA}, N_{CI}$) uniformly distributed across the four regions within the tropical monsoon belt. The spatially averaged total precipitation rates, denoted as  $R$, were calculated for each region during their respective rainy seasons from 1950 to 2022. Taking HI as an example, where the rainy season spans from May to October, the total rainy season precipitation is computed as:
\begin{equation}
R_{HI}(y) = \frac{1}{N_{HI}} \sum_{i=1}^{N_{HI}} \sum_{m=May}^{Oct}\mathcal{R}_{i}^{m,y},\quad y=1950,1951,\ldots,2022,
\label{ave_prate}
\end{equation}
where $y$ indicates the year when the rainy season starts, and $\mathcal{R}_{i}^{m,y}$ is the monthly averaged precipitation rate for node $i$ in month $m$, year $y$.

\subsection{Climate network}
In this study, we construct a series of Climate Networks (CNs) for each calendar year 
$y$ from 1949 to 2021. The strength and direction of links between each pair of nodes in the constructed CNs are determined based on the similarity measure of their respective temperature time series. To account for Earth's spherical geometry, we selected 6,242 latitude-longitude grid points as CN nodes, ensuring an approximately homogeneous global coverage.

\subsubsection{Links}
To define links between each pair of node $i$ and node $j$, we follow \cite{yamasaki2008climate,fan2022network} and compute the time-delayed, cross-correlation function
\begin{equation}
    C^y_{ij}(-\tau) = \frac{\langle T^y_{i}(t)T^y_{j}(t-\tau)\rangle - \langle T^y_{i}(t)\rangle \langle T^y_{j}(t-\tau)\rangle}
    { \sqrt{\langle ( T^y_{i}(t) - \langle T^y_{i}(t) \rangle )^2 \rangle} \sqrt{\langle ( T^y_{j}(t-\tau) - \langle T^y_{j}(t-\tau) \rangle )^2 \rangle}},
\end{equation}
and
\begin{equation}
    C^y_{ij}(\tau) = \frac{\langle T^y_{i}(t-\tau)T^y_{j}(t)\rangle - \langle T^y_{i}(t-\tau)\rangle \langle T^y_{j}(t)\rangle}
    { \sqrt{\langle ( T^y_{i}(t-\tau) - \langle T^y_{i}(t-\tau) \rangle )^2 \rangle} \sqrt{\langle ( T^y_{j}(t) - \langle T^y_{j}(t) \rangle )^2 \rangle}},
\end{equation}
where $< >$ denote the average over the past 365 days, i.e.,
\begin{equation}
    \langle f(t) \rangle = \frac{1}{365}\sum_{k=1}^{365}f(t-k),
\end{equation}
where $t$ stands for the end of time series(i.e., 31 Dec. of the year $y$). Moreover, we define the time lags $\tau \in [0,200]$ to ensure a reliable estimate of the background noise level.

Next, we identify the highest peak ($\max\limits_{\tau}(C^y_{ij}(\tau))$) and lowest valley ($\min\limits_{\tau}(C^y_{ij}(\tau))$) in the cross-correlation function and denote the corresponding time lags as $\theta^{+}_{ij}$ (i.e., $\max(C^y_{ij}(\tau))=C^y(\theta^{+}_{ij})$) and $\theta^{-}_{ij}$ (i.e., $\min(C^y_{ij}(\tau))=C^y(\theta^{-}_{ij})$), respectively. We then define the \textit{positive} and \textit{negative} strengths of the link between pairs of nodes (e.g., node $i$ and node $j$) in the networks as follow:
\begin{equation}
W^{+}_{ij} = \frac{\max(C_{ij}) - {\rm mean}(C_{ij})}{{\rm std}(C_{ij})},
    \label{w_pos}
\end{equation}
\begin{equation}
W^{-}_{ij} = \frac{\min(C_{ij}) - {\rm mean}(C_{ij})}{{\rm std}(C_{ij})},    \label{w_neg}
\end{equation}
where $\max$, $\min$, $\rm mean$ and $\rm std$ are the maximum, minimum, mean and standard deviation of the cross-correlation function, respectively. Additionally, the signs of the time lags indicate the direction of each link. Specifically, a positive time lag ($\theta_{ij} \geq 0$) means the direction of the link is from node $i$ to node $j$, while a negative time lag ($\theta_{ij} < 0$) means the direction is from node $j$ to node $i$. If the time lag is zero, the direction could be either from $i$ to $j$ or from $j$ to $i$. There are two typical time lags corresponding to the positive and negative strengths of each link, and the directions corresponding to these strengths for the same link could be different.

Subsequently, we establish two distinct networks for each calendar year to differentiate between positive and negative link strengths. This dual-network approach allows us to capture both positive and negative interactions between nodes. Each network is characterized by an anti-symmetric adjacency matrix denoted as $\mathbf{A}(y)$. In this matrix, the entry $a_{ij}(y)$ signifies the presence of a link from node $i$ to node $j$ with a value of $1$, while a link from node $j$ to node $i$ is represented by $-1$ for the year $y$.

\subsubsection{Node Degree}
To explore the dynamic aspects of the global climate system related to precipitation within tropical monsoon regions, we analyze the temporal evolution of key Climate Network (CN) properties. Our focus is on the degree centrality of nodes, a fundamental parameter in network theory that quantifies the total number of connected links \cite{albert2002statistical}. Within the positive and negative CNs of a given year $y$, each node has two distinct degree measures: in-degree and out-degree. The in-degree represents the average strength of incoming links, while the out-degree signifies the average strength of outgoing links. Mathematically, these degree measures are defined as:
\begin{equation}
    k^{in\pm}_{i}(y) = \frac{\sum_{j=1,j\neq i}^{N}W^{\pm}_{ij}(y)I_{a_{ij}(y)=-1}}{\sum_{j=1, j\neq i}^{N}I_{a_{ij}(y)=-1}},
    k^{out\pm}_{j}(y) = \frac{\sum_{i=1,i\neq j}^{N}W^{\pm}_{ij}(y)I_{a_{ij}(y)=1}}{\sum_{i=1, i\neq j}^{N}I_{a_{ij}(y)=1}},
    \label{degree}
\end{equation}
where $N$ equals to the total number of nodes(i.e., 6,242) and $I$ is the indicator function. 

%Therefore, we build yearly weighted and averaged $k^{in\pm}$ and $k^{out\pm}$ time series for each node $i$ and denote them as $\mathbf{k^{in\pm}_{i}} \triangleq \{k_{i}^{in\pm}(y):y=1949,1950,...,2021\}$ and $\mathbf{k^{out\pm}_{i}} \triangleq \{k_{i}^{out\pm}(y):y=1949,1950,...,2021\}$, respectively.

\subsection{Linear Models Construction \label{parameter}}
Within our proposed framework, we employ a linear model with an updating learning set to forecast precipitation for a given year $y$. In this approach, we utilize data from the $L$ preceding years to make the prediction, where $L$ ranges from 5 to 40 years. Specifically, for predicting the precipitation rate in year $\eta$, the learning set comprises the $L$ years immediately preceding that target year as follows:
\begin{equation}
    {\bf Y^{\eta}} \triangleq \{R(y): y=\eta-1, \eta-2,...,\eta-L+1, \eta-L\}.
\end{equation}
Moreover, we have a vector of {\it Predictor} corresponding to the precipitation rate vector $\mathbf{Y}^{\eta}$, which is denoted as:
\begin{equation}
    {\bf X^{\eta-1}} \triangleq \{k(y): y=\eta-2, \eta-3,...,\eta-L, \eta-L-1\}.
\end{equation}
Next, to solve the linear model, 
\begin{equation}
    {\bf Y^{\eta}} = \alpha^{\eta}\cdot{\bf X^{\eta-1}} + \beta^{\eta}.
\end{equation}
we employ the Ordinary Least Squares (OLS) regression technique, which provides the optimal linear unbiased estimator of the model parameters. Thus, the predicted value of year $\eta$  can be calculated as:
\begin{equation}
    \tilde{R}(\eta) = \alpha^{\eta}\cdot k(\eta-1) + \beta^{\eta}.
\end{equation}

We evaluate the forecasting skill associated with each specific $L$ using metrics like  Mean Square Error (RMSE) and Mean Absolute Percentage Error (MAPE). RMSE is widely employed in climate and environmental research \cite{willmott2005advantages}. For the predicted values $\tilde{R}(\eta)$ and the observed values $R(\eta)$ of our proposed model, the RMSE is calculated as
\begin{equation}
	RMSE = \sqrt{\frac{1}{n}\sum_{\eta=\eta_1}^{\eta_n}\left( R(\eta)-\tilde{R}(\eta) \right)^2}.
	\label{RMSE}
\end{equation} 
On the other hand, MAPE provides a measure of the relative error and is defined as:
\begin{equation}
	MAPE = \frac{1}{n}\sum_{\eta=\eta_1}^{\eta_n} \left| \frac{R(\eta)-\tilde{R}(\eta)}{R(\eta)} \right|.
	\label{MAPE}
\end{equation}
MAPE is particularly useful in tasks where sensitivity to relative variations is more important than sensitivity to absolute variations. Both RMSE and MAPE have a numerical range of $[0, +\infty)$, with smaller values indicating better predictive performance.

\clearpage
% \section*{Data availability}
% The data represented in Figs. 1–4 are available as Source Data. All other data that support the plots within this paper and other findings of this study are available from the corresponding author upon reasonable request.

\section*{code availability}
The Python codes used for the analysis are available on GitHub (\url{https://github.com/fanjingfang/Monsoon}).

{\section*{Acknowledgements}}
This work was supported by the National Natural Science Foundation of China (Grant No. 42450183, 12275020, 12135003, 12205025, 42461144209) and the Ministry of Science and Technology of China (2023YFE0109000). J.F. is supported by the Fundamental Research Funds for the Central Universities.

{\section*{Author Contributions}
J.M and J.F designed the research. G.R performed the analysis. G.R, J.M and J.F prepared the manuscript, G.R, J.M and J.F discussed results, and contributed to writing the manuscript. J.M and J.F led the writing of the manuscript. }

\section*{Additional information}
Supplementary Information is available in the online version of the paper.

\section*{Competing interests}
The authors declare no competing interests.

\clearpage
\bibliographystyle{naturemag}
\bibliography{Refs}

\begin{thebibliography}{10}
\expandafter\ifx\csname url\endcsname\relax
  \def\url#1{\texttt{#1}}\fi
\expandafter\ifx\csname urlprefix\endcsname\relax\def\urlprefix{URL }\fi
\providecommand{\bibinfo}[2]{#2}
\providecommand{\eprint}[2][]{\url{#2}}

\bibitem{fu1999variation}
\bibinfo{author}{Fu, C.} \& \bibinfo{author}{Wen, G.}
\newblock \bibinfo{title}{Variation of ecosystems over east asia in association
  with seasonal, interannual and decadal monsoon climate variability}.
\newblock \emph{\bibinfo{journal}{Climatic Change}}
  \textbf{\bibinfo{volume}{43}}, \bibinfo{pages}{477--494}
  (\bibinfo{year}{1999}).

\bibitem{wang2006asian}
\bibinfo{author}{Wang, B.}, \bibinfo{author}{Gadgil, S.} \&
  \bibinfo{author}{Rupa~Kumar, K.}
\newblock \bibinfo{title}{The asian monsoon—agriculture and economy}.
\newblock \emph{\bibinfo{journal}{The Asian Monsoon}} \bibinfo{pages}{651--683}
  (\bibinfo{year}{2006}).

\bibitem{gadgil2006indian}
\bibinfo{author}{Gadgil, S.} \& \bibinfo{author}{Gadgil, S.}
\newblock \bibinfo{title}{The indian monsoon, gdp and agriculture}.
\newblock \emph{\bibinfo{journal}{Economic and political weekly}}
  \bibinfo{pages}{4887--4895} (\bibinfo{year}{2006}).

\bibitem{balek1983hydrology}
\bibinfo{author}{Balek, J.}
\newblock \emph{\bibinfo{title}{Hydrology and water resources in tropical
  regions}} (\bibinfo{publisher}{Elsevier}, \bibinfo{year}{1983}).

\bibitem{worden2007importance}
\bibinfo{author}{Worden, J.}, \bibinfo{author}{Noone, D.} \&
  \bibinfo{author}{Bowman, K.}
\newblock \bibinfo{title}{Importance of rain evaporation and continental
  convection in the tropical water cycle}.
\newblock \emph{\bibinfo{journal}{Nature}} \textbf{\bibinfo{volume}{445}},
  \bibinfo{pages}{528--532} (\bibinfo{year}{2007}).

\bibitem{trenberth2000global}
\bibinfo{author}{Trenberth, K.~E.}, \bibinfo{author}{Stepaniak, D.~P.} \&
  \bibinfo{author}{Caron, J.~M.}
\newblock \bibinfo{title}{The global monsoon as seen through the divergent
  atmospheric circulation}.
\newblock \emph{\bibinfo{journal}{Journal of Climate}}
  \textbf{\bibinfo{volume}{13}}, \bibinfo{pages}{3969--3993}
  (\bibinfo{year}{2000}).

\bibitem{zhou2011global}
\bibinfo{author}{Zhou, T.}, \bibinfo{author}{Hsu, H.} \&
  \bibinfo{author}{Matsumoto, J.}
\newblock \bibinfo{title}{The global monsoon system, research and forecast,
  vol. 2} (\bibinfo{year}{2011}).

\bibitem{slingo2011uncertainty}
\bibinfo{author}{Slingo, J.} \& \bibinfo{author}{Palmer, T.}
\newblock \bibinfo{title}{Uncertainty in weather and climate prediction}.
\newblock \emph{\bibinfo{journal}{Philosophical Transactions of the Royal
  Society A: Mathematical, Physical and Engineering Sciences}}
  \textbf{\bibinfo{volume}{369}}, \bibinfo{pages}{4751--4767}
  (\bibinfo{year}{2011}).

\bibitem{collins2013long}
\bibinfo{author}{Collins, M.} \emph{et~al.}
\newblock \bibinfo{title}{Long-term climate change: projections, commitments
  and irreversibility}  (\bibinfo{year}{2013}).

\bibitem{wang2015rethinking}
\bibinfo{author}{Wang, B.} \emph{et~al.}
\newblock \bibinfo{title}{Rethinking indian monsoon rainfall prediction in the
  context of recent global warming}.
\newblock \emph{\bibinfo{journal}{Nature communications}}
  \textbf{\bibinfo{volume}{6}}, \bibinfo{pages}{7154} (\bibinfo{year}{2015}).

\bibitem{kajtar2017tropical}
\bibinfo{author}{Kajtar, J.~B.}, \bibinfo{author}{Santoso, A.},
  \bibinfo{author}{England, M.~H.} \& \bibinfo{author}{Cai, W.}
\newblock \bibinfo{title}{Tropical climate variability: interactions across the
  pacific, indian, and atlantic oceans}.
\newblock \emph{\bibinfo{journal}{Climate Dynamics}}
  \textbf{\bibinfo{volume}{48}}, \bibinfo{pages}{2173--2190}
  (\bibinfo{year}{2017}).

\bibitem{sperber2013asian}
\bibinfo{author}{Sperber, K.} \emph{et~al.}
\newblock \bibinfo{title}{The asian summer monsoon: an intercomparison of cmip5
  vs. cmip3 simulations of the late 20th century}.
\newblock \emph{\bibinfo{journal}{Climate dynamics}}
  \textbf{\bibinfo{volume}{41}}, \bibinfo{pages}{2711--2744}
  (\bibinfo{year}{2013}).

\bibitem{fan2012improving}
\bibinfo{author}{Fan, K.}, \bibinfo{author}{Liu, Y.} \& \bibinfo{author}{Chen,
  H.}
\newblock \bibinfo{title}{Improving the prediction of the east asian summer
  monsoon: New approaches}.
\newblock \emph{\bibinfo{journal}{Weather and Forecasting}}
  \textbf{\bibinfo{volume}{27}}, \bibinfo{pages}{1017--1030}
  (\bibinfo{year}{2012}).

\bibitem{shi2021significant}
\bibinfo{author}{Shi, P.} \emph{et~al.}
\newblock \bibinfo{title}{Significant land contributions to interannual
  predictability of east asian summer monsoon rainfall}.
\newblock \emph{\bibinfo{journal}{Earth's Future}}
  \textbf{\bibinfo{volume}{9}}, \bibinfo{pages}{e2020EF001762}
  (\bibinfo{year}{2021}).

\bibitem{pathak2022fourcastnet}
\bibinfo{author}{Pathak, J.} \emph{et~al.}
\newblock \bibinfo{title}{Fourcastnet: A global data-driven high-resolution
  weather model using adaptive fourier neural operators}.
\newblock \emph{\bibinfo{journal}{arXiv preprint arXiv:2202.11214}}
  (\bibinfo{year}{2022}).

\bibitem{lam2023learning}
\bibinfo{author}{Lam, R.} \emph{et~al.}
\newblock \bibinfo{title}{Learning skillful medium-range global weather
  forecasting}.
\newblock \emph{\bibinfo{journal}{Science}} \bibinfo{pages}{eadi2336}
  (\bibinfo{year}{2023}).

\bibitem{bi2023accurate}
\bibinfo{author}{Bi, K.} \emph{et~al.}
\newblock \bibinfo{title}{Accurate medium-range global weather forecasting with
  3d neural networks}.
\newblock \emph{\bibinfo{journal}{Nature}} \bibinfo{pages}{1--6}
  (\bibinfo{year}{2023}).

\bibitem{zhang2023skilful}
\bibinfo{author}{Zhang, Y.} \emph{et~al.}
\newblock \bibinfo{title}{Skilful nowcasting of extreme precipitation with
  nowcastnet}.
\newblock \emph{\bibinfo{journal}{Nature}} \textbf{\bibinfo{volume}{619}},
  \bibinfo{pages}{526--532} (\bibinfo{year}{2023}).

\bibitem{price2023gencast}
\bibinfo{author}{Price, I.} \emph{et~al.}
\newblock \bibinfo{title}{Gencast: Diffusion-based ensemble forecasting for
  medium-range weather}.
\newblock \emph{\bibinfo{journal}{arXiv preprint arXiv:2312.15796}}
  (\bibinfo{year}{2023}).

\bibitem{watts1998collective}
\bibinfo{author}{Watts, D.~J.} \& \bibinfo{author}{Strogatz, S.~H.}
\newblock \bibinfo{title}{Collective dynamics of ‘small-world’networks}.
\newblock \emph{\bibinfo{journal}{nature}} \textbf{\bibinfo{volume}{393}},
  \bibinfo{pages}{440--442} (\bibinfo{year}{1998}).

\bibitem{albert2002statistical}
\bibinfo{author}{Albert, R.} \& \bibinfo{author}{Barab{\'a}si, A.-L.}
\newblock \bibinfo{title}{Statistical mechanics of complex networks}.
\newblock \emph{\bibinfo{journal}{Reviews of modern physics}}
  \textbf{\bibinfo{volume}{74}}, \bibinfo{pages}{47} (\bibinfo{year}{2002}).

\bibitem{newman2018networks}
\bibinfo{author}{Newman, M.}
\newblock \emph{\bibinfo{title}{Networks}} (\bibinfo{publisher}{Oxford
  university press}, \bibinfo{year}{2018}).

\bibitem{cohen2010complex}
\bibinfo{author}{Cohen, R.} \& \bibinfo{author}{Havlin, S.}
\newblock \emph{\bibinfo{title}{Complex networks: structure, robustness and
  function}} (\bibinfo{publisher}{Cambridge university press},
  \bibinfo{year}{2010}).

\bibitem{tsonis2004architecture}
\bibinfo{author}{Tsonis, A.~A.} \& \bibinfo{author}{Roebber, P.~J.}
\newblock \bibinfo{title}{The architecture of the climate network}.
\newblock \emph{\bibinfo{journal}{Physica A: Statistical Mechanics and its
  Applications}} \textbf{\bibinfo{volume}{333}}, \bibinfo{pages}{497--504}
  (\bibinfo{year}{2004}).

\bibitem{yamasaki2008climate}
\bibinfo{author}{Yamasaki, K.}, \bibinfo{author}{Gozolchiani, A.} \&
  \bibinfo{author}{Havlin, S.}
\newblock \bibinfo{title}{Climate networks around the globe are significantly
  affected by el nino}.
\newblock \emph{\bibinfo{journal}{Physical review letters}}
  \textbf{\bibinfo{volume}{100}}, \bibinfo{pages}{228501}
  (\bibinfo{year}{2008}).

\bibitem{boers_prediction_2014}
\bibinfo{author}{Boers, N.} \emph{et~al.}
\newblock \bibinfo{title}{Prediction of extreme floods in the eastern {Central}
  {Andes} based on a complex networks approach}.
\newblock \emph{\bibinfo{journal}{Nat Commun}} \textbf{\bibinfo{volume}{5}},
  \bibinfo{pages}{1--7} (\bibinfo{year}{2014}).

\bibitem{boers_complex_2019}
\bibinfo{author}{Boers, N.} \emph{et~al.}
\newblock \bibinfo{title}{Complex networks reveal global pattern of
  extreme-rainfall teleconnections}.
\newblock \emph{\bibinfo{journal}{Nature}} \textbf{\bibinfo{volume}{566}},
  \bibinfo{pages}{373--377} (\bibinfo{year}{2019}).

\bibitem{feng_are_2014}
\bibinfo{author}{Feng, Q.~Y.} \& \bibinfo{author}{Dijkstra, H.}
\newblock \bibinfo{title}{Are {North} {Atlantic} multidecadal {SST} anomalies
  westward propagating?}
\newblock \emph{\bibinfo{journal}{Geophysical Research Letters}}
  \textbf{\bibinfo{volume}{41}}, \bibinfo{pages}{541--546}
  (\bibinfo{year}{2014}).

\bibitem{dijkstra2019networks}
\bibinfo{author}{Dijkstra, H.~A.}, \bibinfo{author}{Hern{\'a}ndez-Garc{\'\i}a,
  E.}, \bibinfo{author}{Masoller, C.} \& \bibinfo{author}{Barreiro, M.}
\newblock \emph{\bibinfo{title}{Networks in climate}}
  (\bibinfo{publisher}{Cambridge University Press}, \bibinfo{year}{2019}).

\bibitem{ludescher2021network}
\bibinfo{author}{Ludescher, J.} \emph{et~al.}
\newblock \bibinfo{title}{Network-based forecasting of climate phenomena}.
\newblock \emph{\bibinfo{journal}{Proceedings of the National Academy of
  Sciences}} \textbf{\bibinfo{volume}{118}}, \bibinfo{pages}{e1922872118}
  (\bibinfo{year}{2021}).

\bibitem{ludescher2013improved}
\bibinfo{author}{Ludescher, J.} \emph{et~al.}
\newblock \bibinfo{title}{Improved el ni{\~n}o forecasting by cooperativity
  detection}.
\newblock \emph{\bibinfo{journal}{Proceedings of the National Academy of
  Sciences}} \textbf{\bibinfo{volume}{110}}, \bibinfo{pages}{11742--11745}
  (\bibinfo{year}{2013}).

\bibitem{meng2020complexity}
\bibinfo{author}{Meng, J.} \emph{et~al.}
\newblock \bibinfo{title}{Complexity-based approach for el ni{\~n}o magnitude
  forecasting before the spring predictability barrier}.
\newblock \emph{\bibinfo{journal}{Proceedings of the National Academy of
  Sciences}} \textbf{\bibinfo{volume}{117}}, \bibinfo{pages}{177--183}
  (\bibinfo{year}{2020}).

\bibitem{fan2022network}
\bibinfo{author}{Fan, J.} \emph{et~al.}
\newblock \bibinfo{title}{Network-based approach and climate change benefits
  for forecasting the amount of indian monsoon rainfall}.
\newblock \emph{\bibinfo{journal}{Journal of Climate}}
  \textbf{\bibinfo{volume}{35}}, \bibinfo{pages}{1009--1020}
  (\bibinfo{year}{2022}).

\bibitem{johnson2019seas5}
\bibinfo{author}{Johnson, S.~J.} \emph{et~al.}
\newblock \bibinfo{title}{Seas5: the new ecmwf seasonal forecast system}.
\newblock \emph{\bibinfo{journal}{Geoscientific Model Development}}
  \textbf{\bibinfo{volume}{12}}, \bibinfo{pages}{1087--1117}
  (\bibinfo{year}{2019}).

\bibitem{saha2014ncep}
\bibinfo{author}{Saha, S.} \emph{et~al.}
\newblock \bibinfo{title}{The ncep climate forecast system version 2}.
\newblock \emph{\bibinfo{journal}{Journal of climate}}
  \textbf{\bibinfo{volume}{27}}, \bibinfo{pages}{2185--2208}
  (\bibinfo{year}{2014}).

\bibitem{willmott2005advantages}
\bibinfo{author}{Willmott, C.~J.} \& \bibinfo{author}{Matsuura, K.}
\newblock \bibinfo{title}{Advantages of the mean absolute error (mae) over the
  root mean square error (rmse) in assessing average model performance}.
\newblock \emph{\bibinfo{journal}{Climate research}}
  \textbf{\bibinfo{volume}{30}}, \bibinfo{pages}{79--82}
  (\bibinfo{year}{2005}).

\bibitem{de2016mean}
\bibinfo{author}{De~Myttenaere, A.}, \bibinfo{author}{Golden, B.},
  \bibinfo{author}{Le~Grand, B.} \& \bibinfo{author}{Rossi, F.}
\newblock \bibinfo{title}{Mean absolute percentage error for regression
  models}.
\newblock \emph{\bibinfo{journal}{Neurocomputing}}
  \textbf{\bibinfo{volume}{192}}, \bibinfo{pages}{38--48}
  (\bibinfo{year}{2016}).

\bibitem{cohen2009pearson}
\bibinfo{author}{Cohen, I.} \emph{et~al.}
\newblock \bibinfo{title}{Pearson correlation coefficient}.
\newblock \emph{\bibinfo{journal}{Noise reduction in speech processing}}
  \bibinfo{pages}{1--4} (\bibinfo{year}{2009}).

\bibitem{ferreira2022evaluation}
\bibinfo{author}{Ferreira, G.~W.}, \bibinfo{author}{Reboita, M.~S.} \&
  \bibinfo{author}{Drumond, A.}
\newblock \bibinfo{title}{Evaluation of ecmwf-seas5 seasonal temperature and
  precipitation predictions over south america}.
\newblock \emph{\bibinfo{journal}{Climate}} \textbf{\bibinfo{volume}{10}},
  \bibinfo{pages}{128} (\bibinfo{year}{2022}).

\bibitem{chevuturi2021forecast}
\bibinfo{author}{Chevuturi, A.} \emph{et~al.}
\newblock \bibinfo{title}{Forecast skill of the indian monsoon and its onset in
  the ecmwf seasonal forecasting system 5 (seas5)}.
\newblock \emph{\bibinfo{journal}{Climate Dynamics}}
  \textbf{\bibinfo{volume}{56}}, \bibinfo{pages}{2941--2957}
  (\bibinfo{year}{2021}).

\bibitem{lang2014evaluating}
\bibinfo{author}{Lang, Y.} \emph{et~al.}
\newblock \bibinfo{title}{Evaluating skill of seasonal precipitation and
  temperature predictions of ncep cfsv2 forecasts over 17 hydroclimatic regions
  in china}.
\newblock \emph{\bibinfo{journal}{Journal of Hydrometeorology}}
  \textbf{\bibinfo{volume}{15}}, \bibinfo{pages}{1546--1559}
  (\bibinfo{year}{2014}).

\bibitem{gelbrecht2018phase}
\bibinfo{author}{Gelbrecht, M.}, \bibinfo{author}{Boers, N.} \&
  \bibinfo{author}{Kurths, J.}
\newblock \bibinfo{title}{Phase coherence between precipitation in south
  america and rossby waves}.
\newblock \emph{\bibinfo{journal}{Science advances}}
  \textbf{\bibinfo{volume}{4}}, \bibinfo{pages}{eaau3191}
  (\bibinfo{year}{2018}).

\bibitem{changnon2000nino}
\bibinfo{author}{Changnon, S.~A.}
\newblock \emph{\bibinfo{title}{El Ni{\~n}o 1997-1998: the climate event of the
  century}} (\bibinfo{publisher}{Oxford University Press},
  \bibinfo{year}{2000}).

\bibitem{ronghui1989influence}
\bibinfo{author}{Ronghui, H.} \& \bibinfo{author}{Yifang, W.}
\newblock \bibinfo{title}{The influence of enso on the summer climate change in
  china and its mechanism}.
\newblock \emph{\bibinfo{journal}{Advances in Atmospheric Sciences}}
  \textbf{\bibinfo{volume}{6}}, \bibinfo{pages}{21--32} (\bibinfo{year}{1989}).

\bibitem{yuan20172016}
\bibinfo{author}{Yuan, Y.} \emph{et~al.}
\newblock \bibinfo{title}{The 2016 summer floods in china and associated
  physical mechanisms: A comparison with 1998}.
\newblock \emph{\bibinfo{journal}{Journal of Meteorological Research}}
  \textbf{\bibinfo{volume}{31}}, \bibinfo{pages}{261--277}
  (\bibinfo{year}{2017}).

\bibitem{kripalani2001monsoon}
\bibinfo{author}{Kripalani, R.~H.} \& \bibinfo{author}{Kulkarni, A.}
\newblock \bibinfo{title}{Monsoon rainfall variations and teleconnections over
  south and east asia}.
\newblock \emph{\bibinfo{journal}{International Journal of Climatology: A
  Journal of the Royal Meteorological Society}} \textbf{\bibinfo{volume}{21}},
  \bibinfo{pages}{603--616} (\bibinfo{year}{2001}).

\bibitem{nooteboom_using_2018}
\bibinfo{author}{Nooteboom, P.~D.}, \bibinfo{author}{Feng, Q.~Y.},
  \bibinfo{author}{López, C.}, \bibinfo{author}{Hernández-García, E.} \&
  \bibinfo{author}{Dijkstra, H.~A.}
\newblock \bibinfo{title}{Using network theory and machine learning to predict
  {El} {Niño}}.
\newblock \emph{\bibinfo{journal}{Earth System Dynamics}}
  \textbf{\bibinfo{volume}{9}}, \bibinfo{pages}{969--983}
  (\bibinfo{year}{2018}).

\bibitem{petersik_probabilistic_2020}
\bibinfo{author}{Petersik, P.~J.} \& \bibinfo{author}{Dijkstra, H.~A.}
\newblock \bibinfo{title}{Probabilistic {Forecasting} of {El} {Niño} {Using}
  {Neural} {Network} {Models}}.
\newblock \emph{\bibinfo{journal}{Geophysical Research Letters}}
  \textbf{\bibinfo{volume}{47}}, \bibinfo{pages}{e2019GL086423}
  (\bibinfo{year}{2020}).

\end{thebibliography}
\end{document}